\documentstyle[twoside,fleqn,espcrc2,epsf]{article}

\newcommand{\ainv}{{a^{-1}}}
\newcommand{\be}{\begin{equation}}
\newcommand{\ee}{\end{equation}}

\newcommand{\order}{{\cal O}}
\newcommand{\alphaMS}{\alpha_{\overline{MS}}[M_{Z^0}]}

\newcommand{\AmS}{{\protect\the\textfont2
  A\kern-.1667em\lower.5ex\hbox{M}\kern-.125emS}}

\title{Update on Quarkonium Spectroscopy and $\alpha_{strong}$ from
       NRQCD }

\author{C. T. H. Davies, A. J. Lidsey, P. McCallum,
	\address{Dept. of Physics \& Astronomy,
	University of Glasgow,
	Glasgow G12 8QQ, U.K.}
        K. Hornbostel,
        \address{ Southern Methodist University, Dallas, TX 75275}
        G. P. Lepage,
        \address{Newman Laboratory of Nuclear Studies,
          Cornell University, Ithaca, NY 14853}
        J. Shigemitsu,
        \address{Physics Department,
        The Ohio State University,
        Columbus, Ohio 43210, USA.}
        J. Sloan
        \address{Florida State University, SCRI, Tallahassee, FL 32306}
                        }

\begin{document}

\begin{abstract}
NRQCD results for Upsilon and Charmonium using both dynamical
 and quenched configurations are presented.
  We investigate dependence on the light dynamical quark mass.
Preliminary dynamical ($n_f = 2$)
Charmonium data are combined with quenched results
to extract the strong coupling constant $\alpha_P^{(n_f)}$ for the
physical number of light dynamical quarks, $n_f = 3$.
Good agreement is found with calculations based on the Upsilon system.
We show that a
discrepancy in $\alpha_P^{(n_f=0)}$, found between the Upsilon and Charmonium
systems in the quenched theory, disappears upon extrapolating to the
physical number of flavors. Results for the strong coupling constant
$\alpha_{\overline{MS}}$ are presented
and sources of systematic error investigated.

\end{abstract}

\maketitle

\section{Introduction}

The Nonrelativistic QCD (NRQCD) \cite{LepTh,cornell} approach to heavy
quarks on the lattice has been applied very successfully to the $\Upsilon$
 and Charmonium systems in recent years \cite{bethchris,ups,charm}.
Good agreement is found between lattice simulations and observed quarkonium
levels. These investigations have also enabled us  to determine the
b-quark pole- and $\overline{MS}$-masses \cite{mb} and the strength of the
strong coupling constant, $\alpha_s$ \cite{alpha}.

During the past year we have continued studies of quarkonium systems
 with particular emphasis on refining our
$\alpha_s$ determination. We have :

\begin{enumerate}

\item a new high statistics  $\Upsilon$ study on $16^3 \times 32$
Kogut et al \cite{thanks} configurations with $n_f=0$ at $\beta = 6.0$.

\item increased statistics for $\Upsilon$ studies
on $16^3 \times 32$ HEMCGC \cite{thanks} configurations
with $n_f=2$, $am_q = 0.01$ staggered dynamical quarks at $\beta = 5.6$.

\item new $\Upsilon$ results using HEMCGC dynamical configurations with
$am_q = 0.025$.

\item new ( preliminary) Charmonium data using
$16^3 \times 24$ dynamical MILC
\cite{thanks} configurations ($n_f=2$, $am_q = 0.0125$) at $\beta = 5.145$.

\end{enumerate}

\noindent
This has led to the following main results ,

\begin{itemize}
\item  an $\Upsilon$ spectroscopy update

\item  a look at the $m_q$-dependence of $\alpha_P^{(n_f=2)}$

\item extrapolation of $\alpha_P^{(0)}$, $\alpha_P^{(2)} \rightarrow
 \alpha_P^{(n_f=3)}$ from both $\Upsilon$ and Charmonium. \\
 a $5\sim6$ sigma discrepancy at $n_f=0$ is removed through unquenching and
 extrapolation to the physical $n_f = 3$.

\item establishment of a reliable calculation for $\alpha_{\overline{MS}}$ at
 the $Z^0$ mass.

\end{itemize}

\section{$\Upsilon$ in the Quenched Approximation}

The NRQCD action, details of how heavy quark propagators are evaluated,
the smearing functions and the fitting procedures used are described in
detail in several publications \cite{cornell,ups,charm}.
We will review them briefly.

With $v^2\sim0.1$, the b quarks in the $\Upsilon$ system are quite
nonrelativistic. The splittings between spin-averaged levels are around
500GeV $[\order(M_bv^2)]$, which is much smaller than the $\Upsilon$ mass
$[\order(2M_b)]$, indicating that a systemic expansion of the QCD Hamiltonian
in powers of $v^2$ is appropriate. The continuum action density, correct
through $\order(M_bv^4)$, is broken down according to

\be
{\cal L}_{cont}=\psi^{\dagger}(D_t+H_0^{cont})\psi+\psi^{\dagger}
\delta H^{cont}\psi \label{deltaH}
\ee
\vspace{0.1in}

\noindent
$H_0^{cont}$ and $\delta H^{cont}$ are given explicitly in ref.\cite{ups}, we
give here their lattice counterparts $H_0$ and $\delta H$:

$$
H_0=-\frac{\Delta^{(2)}}{2M_b^0} \qquad\mbox{and}$$

\begin{eqnarray*}
\delta H &\hspace{-0.5em} = &\hspace{-1em}{} -c_1\frac{(\Delta^{(2)})^2}
{8(M_b^0)^3}+c_2
\frac{ig}{8(M_b^0)^2}
({\bf \Delta\cdot E}-{\bf E\cdot\Delta})\\ &&\hspace{-1em}{} -c_3
\frac{g}{8(M_b^0)^2}{\bf \sigma\cdot
(\Delta}\times{\bf E}-{\bf E}\times{\bf \Delta})\\ &&\hspace{-1em}{} -c_4
\frac{g}{2M_b^0}{\bf \sigma
\cdot B}+c_5\frac{a^2\Delta^{(4)}}{24M_b^0}-c_6
\frac{a(\Delta^{(2)})^2}{16n(M_b^0)^2}
\end{eqnarray*}

\noindent
The last two terms in $\delta H$ come from finite lattice spacing corrections
to the lattice Laplacian and lattice time derivative, respectively.

The quark propagators are determined from evolution equations that specify the
propagator value, for $t>0$, in terms of the value on the previous timeslice;

\begin{eqnarray*}
G_1&\hspace{-1em}= &\hspace{-1em}{\left(1-\frac{aH_0}{2n}\right)}^n
U_4^{\dagger}
     {\left(1-\frac{aH_0}{2n}\right)}^n \delta_{x,0}\mbox{,}\\
G_{t+1}&\hspace{-1em}= &\hspace{-1em}{\left(1-\frac{aH_0}{2n}\right)}^n
U_4^{\dagger}
     {\left(1-\frac{aH_0}{2n}\right)}^n(1-a\delta H)G_t
\end{eqnarray*}

The quark propagators are combined with the smearing operator $\Gamma$ to
produce a meson propagator

\begin{eqnarray*}
G_{q\overline{q}}({\bf p},t)&\hspace{-0.5em}=&\hspace{-1em}\sum_{\bf
y_1,y_2}^{}
   \hspace{-0.33em}{\rm Tr}\hspace{-0.16em}\left[G_t^{\dagger}
   ({\bf y_2})\Gamma^{(sk)\dagger}
   ({\bf y_1 - y_2})\tilde{G}_t({\bf y_1})\right]\\
   && \times e^{-i\frac{\bf p}
   {2}\cdot{\bf (y_1+y_2)}}
\end{eqnarray*}

with

$$
\tilde{G}_t({\bf y}) \equiv \sum_{\bf x}G_t({\bf y-x})\Gamma^{(sc)}({\bf x})
  e^{i\frac{\bf p}{2}\cdot {\bf x}}$$

\noindent
where the trace is over spin and color. $\Gamma^{(sc)}$, the smearing at the
source
is distinguished from that at the sink, $\Gamma^{(sk)}$.

Two main fitting procedures were used to extract energies and amplitudes.
For the first procedure, a matrix of correlations with $n_{sc}\mbox{,}n_{sk}
=1,2,3$ for $S$ states and $n_{sc}\mbox{,}n_{sk}=1,2$ for spin singlet $P$
states was fit simultaneously, each correlation being fit to an ansatz of the
form

$$
G_{q\overline{q}}(n_{sc},n_{sk};t) = \sum_{k=1}^{N_{exp}} a(n_{sc},k)
  a^{\ast}(n_{sk},k)e^{-E_kt}$$

\noindent
The second procedure involved fitting a row of smeared-local correlations,
$n_{sc}=1,2,3$ or $n_{sc}=1,2$ with $n_{sk} =$ loc, simultaneously, each
correlation
being fit to an ansatz of the form

$$
G_{q\overline{q}}(n_{sc},{\rm loc};t)= \sum_{k=1}^{N_{exp}}b(n_{sc},k)
  e^{-E_kt}$$

\vspace{0.2in}

A comparison of the dimensionless 1S-1$\overline{{}^3{\rm P}}$ and 1S-2S
$\Upsilon$ splittings, once $\order(a^2)$ gluonic corrections are made
 (using perturbation theory $a\delta E_g \sim 0.0036$ for the 1S and
$\sim 0.0023$ for the 2S level),
with experimental values for these splittings yields the following values
for $\ainv$

\vspace{.1in}

     1S - 1P:  \hspace{0.4in} $\ainv$ = 2.59(6) GeV

     1S - 2S:  \hspace{0.4in} $\ainv$ = 2.34(5) GeV

\vspace{.1in}
\noindent
These results represent an increased statistical accuracy of a factor of 2
over our previous results \cite{ups}.
Taking an average of $\ainv=2.4$ GeV, the updated NRQCD $\Upsilon$
spectrum is shown in figures 1 \& 2 (see section 3).
The agreement between figures 1 \& 2 and their previous counterparts \cite{ups}
is good - a significant shift (2$\sigma$) is only observed for the 3S state.
This may be an indication that previous statistical errors, which included
an element of fitting uncertainty, were overestimates.

\section{$\Upsilon$ with Dynamical Quarks}

We now have twice the statistics on $n_f=2$ configurations as compared to
a year ago \cite{alpha}
and have accumulated 6,400 meson propagators per channel and
per choice of smearing at source and sink.
As in the past we find it crucial to use multi-exponential fits to
several correlations simultaneously, in order to extract ground state and
one or two excited state energies for the S-states and the spin averaged
P-states. We refer to \cite{ups,charm} for details and mention here just
that with higher statistics we are now much more sensitive to systematics.
To give one example,
  the statistical errors of the dimensionless 1S-1P splitting
are of order $\sim 0.005$ which is comparable to
shifts expected from \order($a^2$) corrections to the gluonic action.
The latter have been calculated perturbatively to be $a \delta E_g \sim 0.0057$
 for the 1S and $\sim 0.0034$ for the 2S level.
With several groups
now working with $a^2$ improved gluonic actions one would eventually like to
check $a\delta E_g$ nonperturbatively.  These corrections are crucial when
we check for scaling by comparing $\Upsilon$ simulation results at
different $\beta$ values.

By comparing the dimensionless 1S-1P and 1S-2S $\Upsilon$ splittings with
experiment, one determines $\ainv$.  We find

\vspace{.1in}

     1S - 1P:  \hspace{0.4in} $\ainv$ = 2.44(7) GeV

     1S - 2S:  \hspace{0.4in} $\ainv$ = 2.37(10) GeV

\vspace{.1in}
\noindent
Using an average $\ainv$=2.4GeV we show our updated results for the NRQCD
$\Upsilon$ spectrum in figures 1 \& 2.
Upon focusing on the 1P and 2S levels in figure 1, one notices better agreement
 with experiment in the dynamical theory than in the quenched theory.  This
reflects the fact that $\ainv$'s from the 1S-1P and 1S-2S splittings disagree
with each other (at the $\sim4$-sigma level) in the quenched theory.  As
discussed in reference \cite{alpha} the effect of this discrepancy in
$\alpha_P$ disappears when one extrapolates to the $n_f=3$ theory.  This
remains true also of our new $n_f=0$ and $n_f=2$ data.
A similar phenomenon of different quantities giving different  $\ainv$'s
in the quenched theory, whose effects then disappear for the physical number
of dynamical flavors, will be discussed and explained in section 5
 where we compare $\Upsilon$ and Charmonium results.

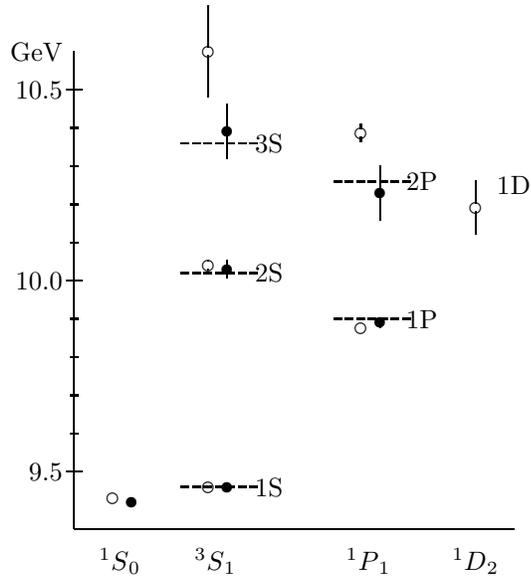
\begin{figure}[t]
\begin{center}
\setlength{\unitlength}{.02in}
\begin{picture}(130,140)(10,930)

\put(15,935){\line(0,1){125}}
\multiput(13,950)(0,50){3}{\line(1,0){4}}
\multiput(14,950)(0,10){10}{\line(1,0){2}}
\put(12,950){\makebox(0,0)[r]{9.5}}
\put(12,1000){\makebox(0,0)[r]{10.0}}
\put(12,1050){\makebox(0,0)[r]{10.5}}
\put(12,1060){\makebox(0,0)[r]{GeV}}
\put(15,935){\line(1,0){115}}



\put(27,930){\makebox(0,0)[t]{${^1S}_0$}}
\put(25,943.1){\circle{3}}
\put(30,942){\circle*{3}}

\put(52,930){\makebox(0,0)[t]{${^3S}_1$}}
\put(66,946){\makebox(0,0){1S}}
\multiput(43,946)(3,0){7}{\line(1,0){2}}
\put(50,946){\circle{3}}
\put(55,946){\circle*{3}}

\put(66,1002){\makebox(0,0){2S}}
\multiput(43,1002)(3,0){7}{\line(1,0){2}}
\put(50,1004.1){\circle{3}}
\put(50,1005.1){\line(0,1){0.2}}
\put(50,1003.1){\line(0,-1){0.2}}
\put(55,1003){\circle*{3}}
\put(55,1004){\line(0,1){1.4}}
\put(55,1002){\line(0,-1){1.4}}

\put(66,1036){\makebox(0,0){3S}}
\multiput(43,1036)(3,0){7}{\line(1,0){2}}
\put(50,1060){\circle{3}}
\put(50,1061){\line(0,1){11}}
\put(50,1059){\line(0,-1){11}}
\put(55,1039.1){\circle*{3}}
\put(55,1039.1){\line(0,1){7.2}}
\put(55,1039.1){\line(0,-1){7.2}}

\put(92,930){\makebox(0,0)[t]{${^1P}_1$}}

\put(106,990){\makebox(0,0){1P}}
\multiput(83,990)(3,0){7}{\line(1,0){2}}
\put(90,987.6){\circle{3}}
\put(95,989){\circle*{3}}
\put(95,990){\line(0,1){0.2}}
\put(95,988){\line(0,-1){0.2}}

\put(106,1026){\makebox(0,0){2P}}
\multiput(83,1026)(3,0){7}{\line(1,0){2}}
\put(90,1038.7){\circle{3}}
\put(90,1039.7){\line(0,1){1.4}}
\put(90,1037.7){\line(0,-1){1.4}}
\put(95,1023){\circle*{3}}
\put(95,1023){\line(0,1){7.2}}
\put(95,1023){\line(0,-1){7.2}}

\put(130,1025){\makebox(0,0){1D}}
\put(120,930){\makebox(0,0)[t]{${^1D}_2$}}
\put(120,1019.2){\circle{3}}
\put(120,1020.2){\line(0,1){6}}
\put(120,1018.2){\line(0,-1){6}}

\end{picture}
\end{center}

\caption{{\bf  $\Upsilon$ Spectrum }
 for $n_f = 0$ (open circles) and $n_f = 2$
(full circles). In both cases, $a^{-1}$ = 2.4GeV was used.
 The dashed horizontal lines denote experimental values.}
\end{figure}

\begin{figure}[bht]
\begin{center}
\setlength{\unitlength}{.02in}
\begin{picture}(100,80)(15,-50)

\put(15,-50){\line(0,1){80}}
\multiput(13,-40)(0,20){4}{\line(1,0){4}}
\multiput(14,-40)(0,10){7}{\line(1,0){2}}
\put(12,-40){\makebox(0,0)[r]{$-40$}}
\put(12,-20){\makebox(0,0)[r]{$-20$}}
\put(12,0){\makebox(0,0)[r]{$0$}}
\put(12,20){\makebox(0,0)[r]{$20$}}
\put(12,30){\makebox(0,0)[r]{MeV}}
\put(15,-50){\line(1,0){100}}


\multiput(28,0)(3,0){7}{\line(1,0){2}}
\put(50,2){\makebox(0,0)[t]{$\Upsilon$}}
\put(35,0){\circle{3}}
\put(40,0){\circle*{3}}

\put(45,-34){\makebox(0,0)[t]{$\eta_b$}}
\put(35,-29.9){\circle{3}}
\put(40,-39){\circle*{3}}

\put(63,-5){\makebox(0,0)[l]{$h_b$}}
\put(70,-0.8){\circle{3}}
\put(75,-2.9){\circle*{3}}
\put(75,-2.9){\line(0,1){1.2}}
\put(75,-2.9){\line(0,-1){1.2}}
\multiput(90,-40)(3,0){7}{\line(1,0){2}}
\put(110,-40){\makebox(0,0)[l]{$\chi_{b0}$}}
\put(97,-24){\circle{3}}
\put(97,-23){\line(0,1){1}}
\put(97,-25){\line(0,-1){1}}
\put(102,-34){\circle*{3}}
\put(102,-34){\line(0,1){5}}
\put(102,-34){\line(0,-1){5}}

\multiput(90,-8)(3,0){7}{\line(1,0){2}}
\put(110,-8){\makebox(0,0)[l]{$\chi_{b1}$}}
\put(97,-8.6){\circle{3}}
\put(102,-7.9){\circle*{3}}
\put(102,-7.9){\line(0,1){2.4}}
\put(102,-7.9){\line(0,-1){2.4}}

\multiput(90,13)(3,0){7}{\line(1,0){2}}
\put(110,13){\makebox(0,0)[l]{$\chi_{b2}$}}
\put(97,10.1){\circle{3}}
\put(102,11.5){\circle*{3}}
\put(102,11.5){\line(0,1){2.4}}
\put(102,11.5){\line(0,-1){2.4}}

\end{picture}
\end{center}
 \caption{{\bf $\Upsilon$ SPIN SPLITTINGS }: Symbols have the
same meanings as in Figure 1.}
\end{figure}
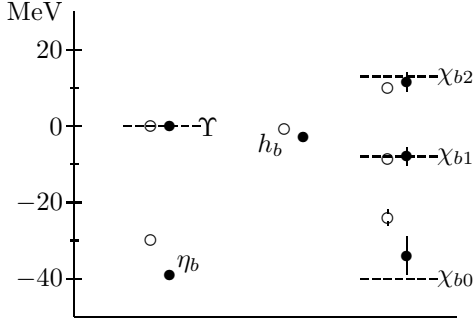

Figure 2 shows spin splittings.  One sees that there are some differences
between experiment and simulations in the
$\chi_b$ splittings.  This is particularly noticeable in the quenched case
and one is tempted to try to extrapolate in $n_f$.  Before carrying out
that exercise, we need to reexamine our fine structure fits.
The fitting procedure used for the P-state fine structure differs from what
was done for the S-states and the $^1P_1$ levels to get the 1S-1P and
1S-2S splittings.  For the latter levels we used multi-exponential
multi-correlation fits as mentioned above.  In order to extract the small
 fine-structure splittings
(these are of order 10 MeV = size of the statistical errors in the $^1P_1$
level)
we looked at ratios between
$^3P_J$ and $^1P_1$ correlations. These were fit to a single exponential.
In past work it was not possible to find a signal for excited state
contributions to these ratios. With our current higher statistics data
it may be possible and necessary to take such contributions into
account.  Further work is required before one can sort out quenching versus
relativistic corrections to the fine structure splittings.

\vspace{.2in}

The last topic to discuss in this section on dynamical $\Upsilon$ data,
concerns dependence on the mass of the dynamical quarks.  The data
discussed above used HEMCGC dynamical configurations with
$n_f=2$ staggered fermions of dimensionless mass $am_q = 0.01$.
  With inverse lattice spacing $a^{-1} \sim 2.4$GeV, this
corresponds to light quarks considerably heavier than the up- or down-
quarks.  This should
not pose a problem as long as $m_q << $ typical momenta in heavy-heavy
systems of about $0.5 \sim 1$GeV (with the higher number
 applicable for $\Upsilon$).  Hence
heavy quark physics with dynamical light quarks should be much less sensitive
to extrapolations in the light quark mass
 than light quark physics.   We have made the first attempt to test this
hypothesis by repeating our $\Upsilon$ runs with $am_q = 0.025$ dynamical
HEMCGC configurations.  The inverse lattice spacings are found to be

\vspace{.1in}

     $am_q = 0.025$

     1S - 1P:  \hspace{0.4in} $\ainv$ = 2.28(15) GeV

     1S - 2S:  \hspace{0.4in} $\ainv$ = 2.17(12) GeV

\vspace{.1in}
\noindent
Comparing with the $\ainv$'s from the $am_q=0.01$ configurations, one finds
a 1-sigma difference for the S-P and a 2-sigma change for the 1S-2S $\ainv$.
It is not clear with present statistics, whether we are observing a true
systematic effect, although the shift downwards in the central values would be
in the right direction for an $am_q$ that has become too large. At given
$\beta$ increasing the dynamical $am_q$ will reduce the difference in $\beta$
between that and a quenched simulation of the same $\ainv$.

An accurate estimate of $\ainv$ is the crucial input for lattice
determinations of the strong coupling constant \cite{alpha,aida}.  So what
one is really interested in here is the $m_q$ dependence of $\alpha_s$
extracted from lattice quarkonium studies.  We use an $\alpha_P$ defined
through the plaquette value,

$$
-lnW_{1,1} = {4\pi \over 3}\alpha_P({3.41 \over a}) [1 - (1.185
+ 0.070n_f)\alpha_P] $$

\noindent
with $\ainv$ from quarkonium splittings setting the scale.  Using the above
 $\ainv$'s and evolving the couplings perturbatively to a common reference
scale of 8.2GeV, one finds

\vspace{.1in}
\noindent
\underline{S-P}

$\alpha_P^{(n_f=2)}[8.2{\rm GeV}] = 0.1793(16)$  \hspace{.1in} $am_q = 0.01$

\vspace{.1in}

$\alpha_P^{(n_f=2)}[8.2{\rm GeV}] = 0.1760(35)$  \hspace{.1in} $am_q = 0.025$

\vspace{.1in}
\noindent
\underline{1S-2S}

$\alpha_P^{(n_f=2)}[8.2{\rm GeV}] = 0.1777(23)$  \hspace{.1in} $am_q = 0.01$

\vspace{.1in}

$\alpha_P^{(n_f=2)}[8.2{\rm GeV}] = 0.1735(28)$  \hspace{.1in} $am_q = 0.025$

\vspace{.1in}
\noindent
One could try extrapolating to $m_q \rightarrow 0$.  Perturbation theory
tells us to extrapolate quadratically in $m_q$. One then finds results very
close to the $am_q = 0.01$ values (0.1799(20) and 0.1785(28) respectively
from S-P and 1S-2S)
More studies of the $m_q$-dependence in dynamical simulations of
quarkonium systems are called for, with higher statistics and
several $m_q$ values.  At the moment our calculations indicate that
the effects of extrapolating from $am_q = 0.01$ down to realistic
light quark masses are very small and less than our statistical errors. Hence,
we
will continue to use the $\ainv$ and $\alpha_P$ values from
the $am_q = 0.01$ simulations for which we have the best statistics.

\section{$\alpha_s$ Determination and Investigation of Systematic Errors}

One of the simplest quantities to calculate in lattice QCD, and to yield a
value for the strong coupling constant, is the expectation value of the
$1 \times 1$ Wilson loop operator. The coupling constant,
$\alpha_P$, is defined through the plaquette value by the perturbative
relation in section 3. With this definition, $\alpha_P$ coincides through
$\order({(\alpha_P)}^2)$ with the coupling $\alpha_V$, defined in refs.
\cite{BLM,LM}.

Using the $\ainv$ values calculated in sections 2 \& 3 ($am_q = 0.01$ results
from section 3) to set the scale and evolving the couplings perturbatively
using the 2-loop formula to the common
reference scale of 8.2 GeV, one obtains

\vspace{0.1in}

\noindent
\underline{1S - 1P}

$\alpha_P^{(n_f=0)}[8.2{\rm GeV}] = 0.1551(11)$

\vspace{0.1in}
$\alpha_P^{(n_f=2)}[8.2{\rm GeV}] = 0.1793(16)$

\vspace{0.1in}
\noindent
\underline{1S - 2S}

$\alpha_P^{(n_f=0)}[8.2{\rm GeV}] = 0.1505(9)$

\vspace{0.1in}
$\alpha_P^{(n_f=2)}[8.2{\rm GeV}] = 0.1777(23)$

\vspace{0.1in}

\noindent
To obtain physical results, one must extrapolate the couplings to $n_f=3$.
The inverse of the coupling $1/\alpha_P^{(n_f)}$ is known to be almost linear
for small changes in $n_f$, hence extrapolating the inverse couplings one
obtains

\vspace{0.1in}
$\alpha_P^{(n_f=3)}[8.2{\rm GeV}] = 0.1945(30)$ \hspace{0.2in} 1S-1P

$\alpha_P^{(n_f=3)}[8.2{\rm GeV}] = 0.1953(43)$ \hspace{0.2in} 1S-2S

\vspace{0.1in}
\noindent
The concordance of these values is more readily appreciated from figure 5.

To enable comparison with other determinations we convert to the
$\overline{MS}$ scheme through the relation

\begin{eqnarray}
\alpha_{\overline{MS}}^{(n_f)}(Q)& =& \alpha_P^{(n_f)}(e^{5/6}Q)\nonumber\\
&& \times \left[1+\frac{2}{\pi}\alpha_P^{(n_f)}+\order({(\alpha_P^{(n_f)})}^2)
  \right]\label{PtoMSbar}
\end{eqnarray}

\noindent
The factor of exp(5/6) absorbs the $n_f$ dependence in (at least) the
$\alpha^2$ term.
Substituting for the $\alpha_P^{(n_f=3)}$ values shown above, evolving
perturbatively to a scale of 1.3GeV, matching from 3 to 4 flavors of
dynamical quarks, evolving once more to 4.3GeV, matching from 4 to 5
flavors of dynamical quarks and finally evolving to the $M_{Z^0}=91.2$GeV
scale, we obtain

\vspace{0.1in}
$\alphaMS = 0.1152(24)$ \hspace{0.2in} 1S-1P

$\alphaMS = 0.1154(26)$ \hspace{0.2in} 1S-2S

\vspace{0.1in}
\noindent
The error in these numbers is dominated by the $\alpha^3$ terms in
equation 2. The central value of course has zero for this coefficient;
one standard deviation allows this coefficient to have magnitude 1.
Ref. \cite{Luscher} has shown that for an $n_f=0$ theory, the coefficient of
the ${(\alpha_P
^{(n_f)})}^3$ term in equation \ref{PtoMSbar} is 0.96, giving  a central
value for $\alphaMS$ of 0.117. The remaining uncertainty would then
be the $n_f$ dependence of the $\alpha^3$ coefficient, and that
may be considerably less than the previous uncertainty.

Although this coefficient is the main source of error, we have also
tested the robustness of our result to other factors.
In one such test,
the $\order(a^2)$ gluonic corrections to the action were omitted, yielding
the results

\vspace{0.1in}
$\alphaMS = 0.1145(24)$ \hspace{0.2in} 1S-1P

$\alphaMS = 0.1154(26)$ \hspace{0.2in} 1S-2S

\vspace{0.1in}
\noindent
The 1S-2S value is identical to that obtained previously with the corrections
in
place since the corrections to the 1S and 2S levels
partially cancel each other. The 1S-1P value is slightly lower
because there are no P-state $\order(a^2)$ gluonic
corrections. So, as expected, although the spectrum
results are accurate enough to
distinguish the presence of this correction our final
value for $\alpha$ is not.

The effect of tadpole-improvement was tested by using quenched results
from \cite{ups} with $u_0 = 1$. We have no dynamical results without
tadpole-improvement so we modified our existing dynamical results to
simulate this situation. We adjusted them by the difference of the
tadpole-improved and tadpole-unimproved quenched energies, since the effect
of unquenching on this difference is a second order effect.
The subsequent $\ainv$ values from the 1S-2S splitting
gave rise to the result

$\alphaMS = 0.1149(27)$ \hspace{0.2in} 1S-2S

\vspace{0.1in}
This result is very close to that obtained with tadpole-improvement
since this has little effect on spin-independent splittings (it is
{\it crucial} for spin splittings).

The effect of the correction terms, $\delta H$,
 in the action of equation\ref{deltaH}
was tested using quenched S-state energies from \cite{ups} with $\delta H = 0$
and the dynamical S-state energies used previously but with a modification
equivalent to that described above for the omission of tadpole-improvement.
 This time
we obtained the result

$\alphaMS = 0.1165(27)$ \hspace{0.2in} 1S-2S

\vspace{0.1in}
\noindent
Although the value for $\alpha_{\overline{MS}}$
 is larger than that for non-zero
$\delta H$, the increase is masked by the
uncertainty incurred when we converted
from $\alpha_P$ to $\alpha_{\overline{MS}}$.
It is clear that higher order relativistic corrections that we do not
include would be completely invisible.

\begin{figure}[t]
\epsfxsize = 7.5cm
\epsfbox[40 30 540 530]{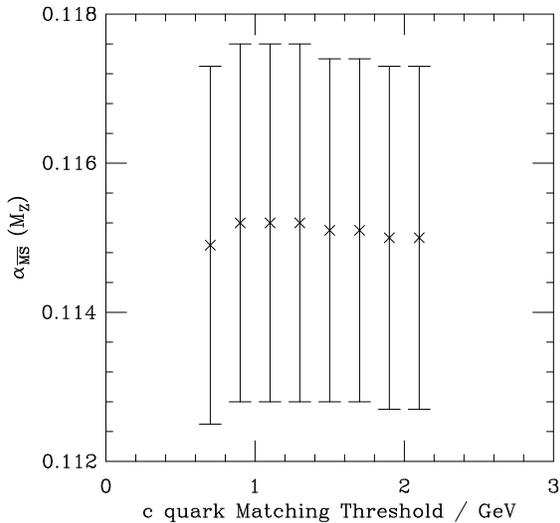}
\vspace{-0.2in}
\caption{Variation of $\alpha_{\overline{MS}}$ with threshold at which
matching from 3 to 4 dynamical quark flavors takes place.}
\label{lowthresh}
\end{figure}

The final factors we investigated were the
thresholds at which we matched from 3 to
4
and from 4 to 5 flavors of dynamical quarks \cite{Rodrigo}. Figure 3 shows
values of
$\alpha_{\overline{MS}}$ obtained from the 1S-1P splitting where the lower
matching
threshold (3 to 4 flavors) was varied around the 1.3GeV value while the upper
matching threshold (4 to 5 flavors) was fixed at 4.3GeV. Figure 4 is similar;
this
time the upper matching threshold was varied around 4.3GeV with the lower fixed
at
1.3GeV. For simplicity we used the values from ref. \cite{Rodrigo} of 1.3 GeV
and 4.3 GeV for
the quark masses, $m_q(m_q)$.
As can be seen from the figures there are comfortable ranges of at least
1.5GeV where there is essentially no effect on our final result, in accord
with the findings of ref. \cite{Rodrigo}.

\begin{figure}[t]
\epsfxsize = 7.5cm
\epsfbox[45 30 520 505]{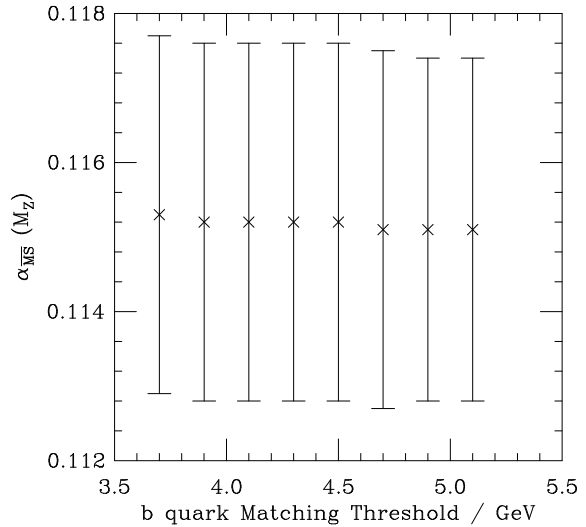}
\vspace{-0.2in}
\caption{Variation of $\alpha_{\overline{MS}}$ with threshold at which
matching from 4 to 5 dynamical quark flavors takes place.}
\label{highthresh}
\end{figure}

\section{Charmonium with Dynamical Quarks and a New Estimate of $\alpha_s$}

The $\ainv$ that feeds into
lattice determinations of $\alpha_s$ can, in principle, be obtained from a
variety of physical quantities.  The reason level splittings in the
$\Upsilon$ system are optimal, is that statistical errors can be
reduced rather efficiently in these systems, and because systematic
errors are also under good control.  Another similar system that can be used
for
 $\alpha_s$ determinations is the Charmonium system \cite{aida}.
  Our collaboration
has studied Charmonium in the quenched approximation using NRQCD heavy
fermions on $\beta = 5.7$ UKQCD configurations.
These results are covered by Christine Davies
at this conference \cite{charm}. We now also have preliminary
 dynamical Charmonium data on $n_f=2$ MILC \cite{thanks}
configurations at $\beta = 5.145$.
Using the Charmonium spin averaged S-P splitting to determine $\ainv$, one
ends up with a value for $\alpha_P^{(n_f=2)}$ of,

\vspace{.1in}

$\alpha_P^{(n_f=2)}[8.2{\rm GeV}] = 0.1758(36)$  \hspace{.2in} Charm

\vspace{.1in}
\noindent
which should be compared with the $\Upsilon$ value
of 0.1793(16) from the previous section.

\vspace{.1in}
\noindent
$\alpha_P^{(n_f)}$ need not agree between Charmonium and $\Upsilon$
 for $n_f = 0$ and $n_f = 2$.  However there should be
agreement for the
physical number of dynamical flavors, $n_f = 3$.
In fact based on the quenched results in section 4 and those discussed by
C.Davies
\cite{charm} one finds for $n_f=0$

\vspace{.1in}

$\alpha_P^{(n_f=0)}[8.2{\rm GeV}] = 0.1480(13)$  \hspace{.2in} Charm

$\alpha_P^{(n_f=0)}[8.2{\rm GeV}] = 0.1551(11)$  \hspace{.2in} $\Upsilon$

\vspace{.1in}
\noindent
We believe the $5 \sim 6$ sigma discrepancy
between Charmonium and $\Upsilon$ is largely a quenching effect combined with
the fact that the two quarkonium systems have different characteristic
energy scales $q^*_{\Upsilon}$ and $q^*_C$ ($q^*_{\Upsilon} > q^*_C$).
In a quenched theory $\alpha(q)$ will run incorrectly between the two $q^*$'s.
 As a consequence the $\Upsilon$ S-P splitting will be underestimated relative
to the same splitting in Charmonium.  Upon comparing with real experimental
data the $\Upsilon$ simulation results will lead to a larger $\ainv$ than
in Charmonium and hence also to a larger $\alpha_P[8.2GeV]$.

  The $n_f=2$ $\alpha_P^{(n_f=2)}$'s agree better between Charmonium and
$\Upsilon$, although one does not expect complete agreement until one has
extrapolated to the physical number of flavors $n_f=3$.  We extrapolate
in $1/\alpha_P$  and obtain,

\vspace{.1in}

$\alpha_P^{(n_f=3)}[8.2{\rm GeV}] = 0.1940(67)$  \hspace{.2in} Charm

$\alpha_P^{(n_f=3)}[8.2{\rm GeV}] = 0.1945(30)$  \hspace{.2in} $\Upsilon$

\vspace{.1in}
\noindent
So, extrapolation to the correct number of flavors has removed the
discrepancy altogether.  The $n_f$ dependence is summarized in figure 5.
The errors on the Charmonium results are still
large, but we are working towards improving them.
The $\alpha_P^{(n_f=3)}$ from Charmonium and the updated value from $\Upsilon$
agree well with our previous published value \cite{alpha}.


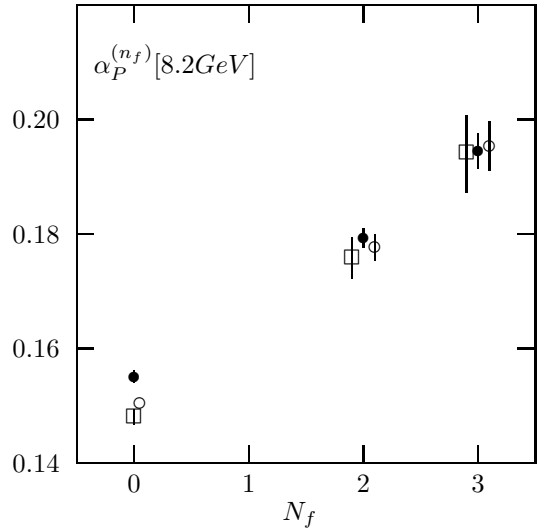
\begin{figure}[t]
\begin{center}
\setlength{\unitlength}{.03in}
\begin{picture}(100,100)(0,135)
\put(10,140){\line(0,1){80}}
\multiput(10,160)(0,20){3}{\line(1,0){3}}
\multiput(90,160)(0,20){3}{\line(-1,0){3}}
\put(7,140){\makebox(0,0)[r]{0.14}}
\put(7,160){\makebox(0,0)[r]{0.16}}
\put(7,180){\makebox(0,0)[r]{0.18}}
\put(7,200){\makebox(0,0)[r]{0.20}}

\put(10,140){\line(1,0){80}}
\put(10,220){\line(1,0){80}}
\put(90,140){\line(0,1){80}}

\multiput(20,140)(20,0){4}{\line(0,1){3}}
\multiput(20,220)(20,0){4}{\line(0,-1){3}}

\put(13,210){\makebox(0,0)[l]{$\alpha_P^{(n_f)}[8.2GeV] $}}
\put(49,130){\makebox(0,0)[b]{$N_f$}}

\put(20,135){\makebox(0,0)[b]{0}}
\put(40,135){\makebox(0,0)[b]{1}}
\put(60,135){\makebox(0,0)[b]{2}}
\put(80,135){\makebox(0,0)[b]{3}}

\put(20,155.1){\circle*{2}}
\put(20,155.1){\line(0,1){1.1}}
\put(20,155.1){\line(0,-1){1.1}}
\put(21,150.5){\circle{2}}
\put(20,147.7){\makebox(0,0){$\Box$}}
\put(20,148.){\line(0,1){1.3}}
\put(20,148.){\line(0,-1){1.3}}

\put(60,179.3){\circle*{2}}
\put(60,179.3){\line(0,1){1.6}}
\put(60,179.3){\line(0,-1){1.6}}
\put(62,177.7){\circle{2}}
\put(62,177.7){\line(0,1){2.3}}
\put(62,177.7){\line(0,-1){2.3}}
\put(58,175.5){\makebox(0,0){$\Box$}}
\put(58,175.8){\line(0,1){3.6}}
\put(58,175.8){\line(0,-1){3.6}}

\put(80,194.5){\circle*{2}}
\put(80,194.5){\line(0,1){3.}}
\put(80,194.5){\line(0,-1){3.}}
\put(82,195.4){\circle{2}}
\put(82,195.4){\line(0,1){4.3}}
\put(82,195.4){\line(0,-1){4.3}}
\put(78,193.7){\makebox(0,0){$\Box$}}
\put(78,194.0){\line(0,1){6.7}}
\put(78,194.0){\line(0,-1){6.7}}

\end{picture}
\end{center}
\caption{  $\alpha_P^{(n_f)}[8.2GeV]$ versus $N_f$ with the scale set by
the $\Upsilon$ 1S-1P
 (full circles), the $\Upsilon$ 1S-2S
(open circles) or the Charmonium 1S-1P (boxes) splitting.}

\end{figure}

In figure 5 we also show $\alpha_P^{(n_f)}$ from $\Upsilon$ 1S-2S splittings
(the open circles). Again a discrepancy at $n_f=0$ disappears upon
going to the $n_f=3$ theory. Arguing why, in the same quarkonium system, one
expects a larger $\alpha_P^{(n_f=0)}$ from a 1S-1P determination than from
the 1S-2S splitting, is somewhat subtle. It no longer suffices to talk
about just one typical characteristic $q^*_{\Upsilon}$ as we did when
comparing $\Upsilon$ with Charmonium. One must look more carefully into
the details of individual binding energies for the S- and P-states and
the $\alpha(q)$ values associated with them. As discussed in reference
 \cite{alpha}, in a theory with incorrect running of $\alpha$ (a running
that is too rapid) the 1S-1P splitting will again be underestimated relative
to the 1S-2S splitting, leading to the phenomenon observed in figure 5. The
important point  here is that we have now observed subtle quenching effects
in several ways and that in each case such effects are removed by including
the appropriate number of dynamical light quarks.

\section{Summary}

We have carried out several tests/updates of lattice determinations of
$\alpha_s$ using quenched and dynamical Quarkonium simulations.
These include investigations of the $m_q$-dependence, determination of
$\alpha_P^{(n_f=3)}$ from  systems and/or splittings
 with different characteristic
scales and investigation into the sources of systematic error of the $\alphaMS$
calculation.
    These tests uphold previous determinations of
$\alpha_P^{(3)}$ and
$\alpha_{\overline{MS}}$ and
give us further confidence in the reliability of those results.
Our final value for $\alphaMS$ remains 0.115(2) if the $\alpha^3$ coefficient
in the matching to $\overline{MS}$ is set to zero, and 0.117 if this
coefficient takes its $n_f$ = 0 value of 0.96.

\vspace{.1in}
\noindent
This work is supported in part by the U.S. DOE, the NSF and by the UK PPARC.
 The numerical computations were carried out at NERSC, the Ohio
Supercomputer Center and the Atlas Centre. We thank the HEMCGC and MILC
collaborations for providing the dynamical configurations and UKQCD,
Kogut et al. and Kilcup et al. for providing the quenched configurations
for these studies.

\end{document}